\documentclass[aps,amsmath,amssymb,reprint,twocolumn]{revtex4-2}
\usepackage{amsmath}
\usepackage{amssymb}
\usepackage{graphicx}
\usepackage{dcolumn}
\usepackage{bm}
\usepackage{hyperref}
\usepackage[misc]{ifsym}
\hypersetup{colorlinks=true,citecolor=blue,,urlcolor=blue,linkcolor=blue}
\graphicspath{{fig/}}
\begin{document}

\title{Finite size effects on hinge states in three-dimensional second-order topological insulators}

\author{Penglei Wang$^{1}$, Juntao Song$^{2}$}
\author{Yong-Lian Zou$^{1}$}
\email{zylian-zylian@163.com}
\affiliation{
$^1$Hunan Key Laboratory for Micro-Nano Energy Materials and Devices and School of
	Physics and Optoelectronics, Xiangtan University, Hunan 411105, China}

\affiliation{
$^2$School of Physical Science and Information Engineering, Hebei Normal University, Hebei 050000, China}

\date{\today}
\begin{abstract}
We investigate the finite size effects of a three-dimensional second-order topological insulator with fourfold rotational symmetry and time-reversal symmetry. Starting from the effective Hamiltonian of the three-dimensional second-order topological insulator, we derive the effective Hamiltonian of four two-dimensional surface states with gaps derived by perturbative methods. Then, the sign alternation of the mass term of the effective Hamiltonian on the adjacent surface leads to the hinge state. In addition, we obtain the effective Hamiltonian and its wave function of one-dimensional gapless hinge states with semi-infinite boundary conditions based on the effective Hamiltonian of two-dimensional surface states. In particular, we find that the hinge states on the two sides of the same surface can couple to produce a finite energy gap.
\end{abstract}

\maketitle

\section{Introduction}
Topological insulators (TIs) are energy band insulators with edge states or surface states protected by topology, and they have attracted great attention due to their exotic properties. A well-known paradigm for TI is the quantum Hall effect with an insulating bulk, but supporting transport of electrons along its surface without backscattering and dissipation, unaffected by defects or impurities. The non-trivial topological states of TI are characterized by topological invariants. The standard paradigm of the TI claims that a D-dimensional TI supports the gapless boundary mode of codimension one ($d_{c}=1$)\cite{1,2,3,4,5,6}. For example, the quantum Hall effect and the quantum spin Hall effect (QSH) support one-dimensional boundary modes, and the three-dimensional topological insulator supports the two-dimensional Dirac surface state\cite{7,8}. A low-energy effective model of HgTe/CdTe quantum well system was established  based on $k\cdot p$ perturbation method, and its gapless states have been confirmed by the experiment with the observation of ballistic edge channels\cite{9}. In recent years, TIs have been extended to high-order topological insulators (HOTI), which  generalized the bulk-boundary correspondence principle. The new paradigm is an n-order D-dimensional TI supporting boundary states of codimension $d_{c}=n$, while the surface states remain gapped\cite{10,11,12}. In this way, the previous referred TIs are first-order topological insulator (FOTI). Three-dimensional “first-order” strong TI supports two-dimensional gapless surface states, while second-order and third-order 3D TIs support one-dimensional hinge modes and zero-dimensional corner modes, respectively\cite{13,14,15,16}. The difference between corner modes and hinge modes is that one-dimensional gapless hinge modes have a spectral flow between insulator the conduction band and valence band, while corner modes do not\cite{12}. The topological properties of HOTIs are protected by symmetries involving spatial transformations, perhaps augmented by time-reversal\cite{17}. These hinge states have been predicted and observed in bismuth crystals\cite{18}, and the resulting helical hinge state can be used for non-consuming electron transport.\\

In HOTIs, hinge states or corner states appear at the boundaries of the system, which requires the system to be finite-sized. The 2D FOTI in Hg/CdTe quantum well system with a finite stripe of width has been studied by Zhou \emph{et al.} \cite{19}. Their study showed that edge states on opposite sides of a QSH model with finite width can couple and create an energy gap. In finite-size HOTI, we expect a similar effect to occur. In this paper, we present an analytical study of the finite size effect of 3DSOTI. We first present the effective Hamiltonian for the gapped surface and study the corresponding energy spectrum for the 2D gapped surface states. Based on these results, we then study 1D gapless hinge states and obtain their effective Hamiltonians. In particular, we study the finite-size effect of hinge states and show that two hinge states on the same surface are coupled to create an energy gap.

\section{Low-energy surface Hamiltionian}
A clean three-dimensional second-order topological insulator (3DSOTI) can be modelled by the following Hamiltonian in momentum space\cite{11}
\begin{equation}
	\begin{aligned}
	H(k)=& [m-t_{i}\sum_{i} \cos (k_{i}a)]\tau_{0}\sigma_{z}s_{0}+A_{i}\sum_{i}sin (k_{i}a)\tau_{0}\sigma_{x}s_{i}\\
		 & +\varLambda[\cos (k_{x}a)-\cos (k_{y}a) ]\tau_{y}\sigma_{y}s_{0}
	\end{aligned}
\end{equation}
where a is the lattice constant. Here $ \tau_{i} $ and $ \sigma_{i} $$ (i=x,y,z) $ are Pauli matrices representing the orbital degrees of freedom, and $ s_{i} $ representing the spin degree. $ m $, $ t_{i} $, $ A_{i} $, $ \varLambda $ are the hopping parameters, and $ 1<|\frac{m}{t}|<3 $\cite{12}. For simplicity, we take $t=t_{x}=t_{y}=t_{z}$ and $ A=A_{x}=A_{y}=A_{z} $. Hopping energy $ t $ is chosen as the energy unit. For $ \varLambda\neq 0 $, the Hamiltonian breaks the time reversal symmetry and the fourfold rotation symmetry respectively, but it is protected by the operator $ R_{4}^{z}T $($ T $ is the time reversal operator and $ R_{4}^{z} $ is the quadruple rotation operator). This model can be seen as a well-known 3DTI with the last term opening gaps with alternating signs if the four surfaces on the (100) and (010) surfaces.\\

We first derive the low-energy effective Hamiltonian for the four surface case. Taking a continuum limit about $ (k_{x},k_{y}, k_{z}) = (0, 0, 0) $, the Hamiltonian becomes
\begin{equation}
	\begin{aligned}
   H=&[m-3t+\dfrac{t}{2}(k_{x}^{2}+k_{y}^{2}+k_{z}^{2})]\tau_{0}\sigma_{z}s_{0}+Ak_{x}\tau_{0}\sigma_{x}s_{x}+\\
     &Ak_{y}\tau_{0}\sigma_{x}s_{y}+Ak_{z}\tau_{0}\sigma_{x}s_{z}-\dfrac{\varLambda}{2}(k_{x}^{2}-k_{y}^{2})\tau_{y}\sigma_{y}s_{0}
	\end{aligned}
\end{equation}
Considering a semi-infinite system for $ x\geqslant0 $, the translational symmetry along the $ x $ direction is destroyed, replacing $ k_{x} \rightarrow-i\partial_{x} $ and splitting the Hamilitonian into two parts, $ H=H_{0}+H_{1} $ with
\begin{equation}
	H_{0}(-i\partial_{x}, k_{y}, k_{z})=(m-3t-\dfrac{t}{2}\partial_{x}^{2})\tau_{0}\sigma_{z}s_{0}-i\partial_{x}\tau_{0}\sigma_{x}s_{x}
\end{equation}
and
\begin{equation}
	\begin{aligned}
	H_{1}(-i\partial_{x}, k_{y}, k_{z})=& (\dfrac{t}{2}k_{y}^{2}+\dfrac{t}{2}k_{z}^{2})\tau_{0}\sigma_{z}s_{0}+Ak_{y}\tau_{0}\sigma_{x}s_{y}+\\
	& Ak_{z}\tau_{0}\sigma_{x}s_{z}+\dfrac{\varLambda}{2}(\partial_{x}^{2}+k_{y}^{2})\tau_{y}\sigma_{y}s_{z}
\end{aligned}
\end{equation}
we assume  $ \varLambda $ is sail, so $ H_{1} $ can be treated as a perturbation. The eigenequation $ H_{0}(-i\partial_{x}, k_{y}, k_{z})\psi_{\alpha}(x)=E\psi_{\alpha}(x) $ is solved under boundary condition $ \psi(x=0)=\psi(x\rightarrow +\infty) =0 $, and four zero-energy solutions are obtained, which are in the form of
\begin{equation}
	\psi_{\alpha}(x)=C_{x}\sin(r_{1}x)e^{-r_{2}x}e^{ik_{y}y}e^{ik_{z}z}X_{\alpha}
\end{equation}
with normalization given by $ |C_{x}^{2}|=4r_{2}(r_{1}^{2}+r_{2}^{2})/r_{1}^{2} $, and $ r_{1}=\sqrt{-\frac{2(m-3t)}{t}-\frac{A^{2}}{t^{2}}} $, $ r_{2}=\frac{A}{t} $. The matrix equation that results from solving (3) can be simplified to $(I-\tau_{0}\otimes\sigma_{y}\otimes s_{x})X_{\alpha}=0$, from which follows that $ X_{\alpha} $ is a positive eigenstate of $\tau_{0}\otimes\sigma_{y}\otimes s_{x}$, here we choose
\begin{equation}
\begin{aligned}
	X_{1} &=|\tau_{z}=+1\rangle \otimes|\sigma_{y}=+1\rangle \otimes |s_{x}=+1\rangle \\
	X_{2} &=|\tau_{z}=+1\rangle \otimes|\sigma_{y}=-1\rangle \otimes |s_{x}=-1\rangle \\
	X_{3} &=|\tau_{z}=-1\rangle \otimes|\sigma_{y}=+1\rangle \otimes |s_{x}=-1\rangle \\
	X_{4} &=|\tau_{z}=-1\rangle \otimes|\sigma_{y}=-1\rangle \otimes |s_{x}=+1\rangle 
\end{aligned}
\end{equation}
These four orthogonal wave functions constitute the basis of surface states. We now project $ H_{1} $ into the subspace spanned by these bases 
\begin{equation}
		H_{eff}^{s}(-i\partial_{x},k_{y}, k_{z})=\int_{0}^{+\infty}dx\psi_{\alpha}^{*}(x)H_{1}(-i\partial_{x}, k_{y}, k_{z})\psi_{\beta}(x)
\end{equation}
therefore, the final form of the effective Hamiltonian is
\begin{equation}
	\begin{aligned}
	H_{eff}^{s}(k_{y}, k_{z})=&-Ak_{y}\mu_{0}\otimes\mu_{z}+Ak_{z}\mu_{0}\otimes\mu_{y}\\
	&+\dfrac{\varLambda(m-3t)}{t}\mu_{y}\otimes\mu_{x}+\dfrac{\varLambda}{2}k_{y}^{2}\mu_{y}\otimes\mu_{x}
\end{aligned}
\end{equation}
where $ \mu_{i} $ are Pauli matrices in the basis $ (\psi_{1}(x),\psi_{2}(x),\psi_{3}(x),\psi_{4}(x)) $.The energy spectrum of the surface state is
\begin{equation}
	E_{\pm}=\pm\frac{1}{2}\sqrt{(\dfrac{2\varLambda(m-3t)}{t}-\varLambda k_{y}^{2})^{2}+4A^{2}(k_{y}^{2}+k_{z}^{2})} 
\end{equation}
It can be seen that as long as $ \varLambda\neq0 $, this surface state has an energy gap.\\

Similarly, the low-energy effective Hamiltonians for the additional three edges are\cite{13}
\begin{equation}
	\begin{aligned}
	H_{eff}^{s}( k_{x}, k_{z}) =&Ak_{x}\mu_{0}\otimes\mu_{z}+Ak_{z}\mu_{z}\otimes\mu_{x}\\
	&-\dfrac{\varLambda(m-3t)}{t}\mu_{y}\otimes\mu_{x}-\dfrac{\varLambda}{2}k_{x}^{2}\mu_{y}\otimes\mu_{x}\\
	H_{eff}^{s}(k_{y}, k_{z})=&Ak_{y}\mu_{0}\otimes\mu_{z}-Ak_{z}\mu_{0}\otimes\mu_{y}\\
	&+\dfrac{\varLambda(m-3t)}{t}\mu_{y}\otimes\mu_{x}+\dfrac{\varLambda}{2}k_{y}^{2}\mu_{y}\otimes\mu_{x}\\
	H_{eff}^{s}( k_{x}, k_{z}) =&-Ak_{x}\mu_{0}\otimes\mu_{z}-Ak_{z}\mu_{z}\otimes\mu_{x}\\
	&-\dfrac{\varLambda(m-3t)}{t}\mu_{y}\otimes\mu_{x}-\dfrac{\varLambda}{2}k_{x}^{2}\mu_{y}\otimes\mu_{x}
\end{aligned}
\end{equation}
The Dirac mass terms on the four surfaces are $ \frac{\varLambda(m-3t)}{t}, -\frac{\varLambda(m-3t)}{t}, \frac{\varLambda(m-3t)}{t}, -\frac{\varLambda(m-3t)}{t} $. It can be seen that the four mass terms have alternating signs on the neighboring surfaces at the four hinges, respectively.
\section{Low-energy Hamiltionian of helical hinge states}
According to Jackiw-Rebbi theory\cite{20}, such a domain wall will bind a one-dimensional zero energy state. Thus, we obtain four hinge states. Based on the Hamiltonian of the four surfaces, hinge state formed at $ x=0 $ and $ y=0 $, here, at the other three states are derived similarly. 
We focus on the surface state of $ x=0 $, if the boundary condition is opened in the $ y $ direction and the periodic boundary condition is maintained in the $ z $ direction so that $ k_{z} $ is still a good quantum number. We replace $ k_{y} $ by $ -i\partial_{y} $ in the Hamiltonian of $ x=0 $ surface, and this Hamiltonian can be written as a sum of two parts $ H^{s}=h_{0}+h_{1} $, in which
\begin{equation}
	h_{0}(-i\partial_{y},k_{z})=iA\partial_{y}\mu_{0}\otimes\mu_{z}-\dfrac{\varLambda}{2}\partial_{y}^{2}\mu_{y}\otimes\mu_{x}+\dfrac{\varLambda(m-3t)}{t}\mu_{y}\otimes\mu_{x}
\end{equation}
and
\begin{equation}
	h_{1}(-i\partial_{y},k_{z})=Ak_{z}\mu_{0}\otimes\mu_{y}
\end{equation}
Solving the eigenvalue equation $ h_{0}\psi_{\alpha}(y)=E\psi_{\alpha}(y) $ under the boundary condition $ \Psi(x,y)=\Psi(0,0)=\Psi(0,\infty)=0 $, we find two zero-energy solutions, whose forms are
\begin{equation}
	\begin{aligned}
			\psi_{1}(y)=C_{y}\sin(q_{1}y)e^{-q_{2}y}e^{ik_{z}z}Y_{1}\\				\psi_{2}(y)=C_{y}\sin(q_{1}y)e^{-q_{2}y}e^{ik_{z}z}Y_{2}
	\end{aligned}
\end{equation}
with normalization given by $ |C_{y}^{2}|=4q_{2}(q_{1}^{2}+q_{2}^{2})/q_{1}^{2} $, and $ q_{1}=\sqrt{-\frac{A^{2}}{\varLambda^{2}}-\frac{2(m-3t)}{t}} $, $ q_{2}=\frac{A}{\varLambda} $.  The matrix equation that results from solving (5) can be simplified to $(I-\mu_{y}\otimes\mu_{y})Y_{1,2}=0$, from which follows that $ Y_{1,2} $ is a positive eigenstate of $\mu_{y}\otimes\mu_{y}$, i.e., $ Y_{1}=(i,0,0,1) $ or $ Y_{2}=(0,i,1,0) $. 
These two orthogonal wave functions construct the basis of the hinge states. We now project the  Hamiltonian $ h_{1} $ into the subspace spanned by these two states to find the low-energy Hamiltonian of the $ (x,y)=(0,0) $ hinge 
\begin{equation}
		H_{eff}^{h}(-i\partial_{y}, k_{z})=\int_{0}^{+\infty}dy\psi_{\alpha}^{*}(y)h_{1}(-i\partial_{y}, k_{z})\psi_{\beta}(y)
\end{equation}
therefore, the final form effective Hamiltonian of the hinge states is
\begin{equation}
	H_{eff}^{h}(k_{z})=-Ak_{z}\eta_{y}
\end{equation}
where $ \eta_{y} $ are Pauli matrices in the basis $ (\psi_{1}(y),\psi_{2}(y)) $, and the energy spectrum of the boundary state is
\begin{equation}
	E_{\pm}=\pm Ak_{z}
\end{equation}
\section{Finite size effects of hinge states}
In the previous section, we obtained the Hamiltonian for a single hinge state in a semi-infinite region. Because of the finite size of the system, the four hinge states will hybridize with each other on the same surface. For simplicity, we consider the hybridization of two hinge states on a surface and neglect the impact of the other two hinge states. In this section, we solve the energy spectrum and wave function of two hinge states in the $ x=0 $ surface under finite size. To have solutions for the energy spectrum and wave functions, we solve the Schr\"{o}dinger equations for the Hamiltonian(8)
\begin{equation}
	H^{s}\psi(y)=E\psi(y)
\end{equation}
\begin{figure}[htbp]
	\centering
	\includegraphics[width=8cm]{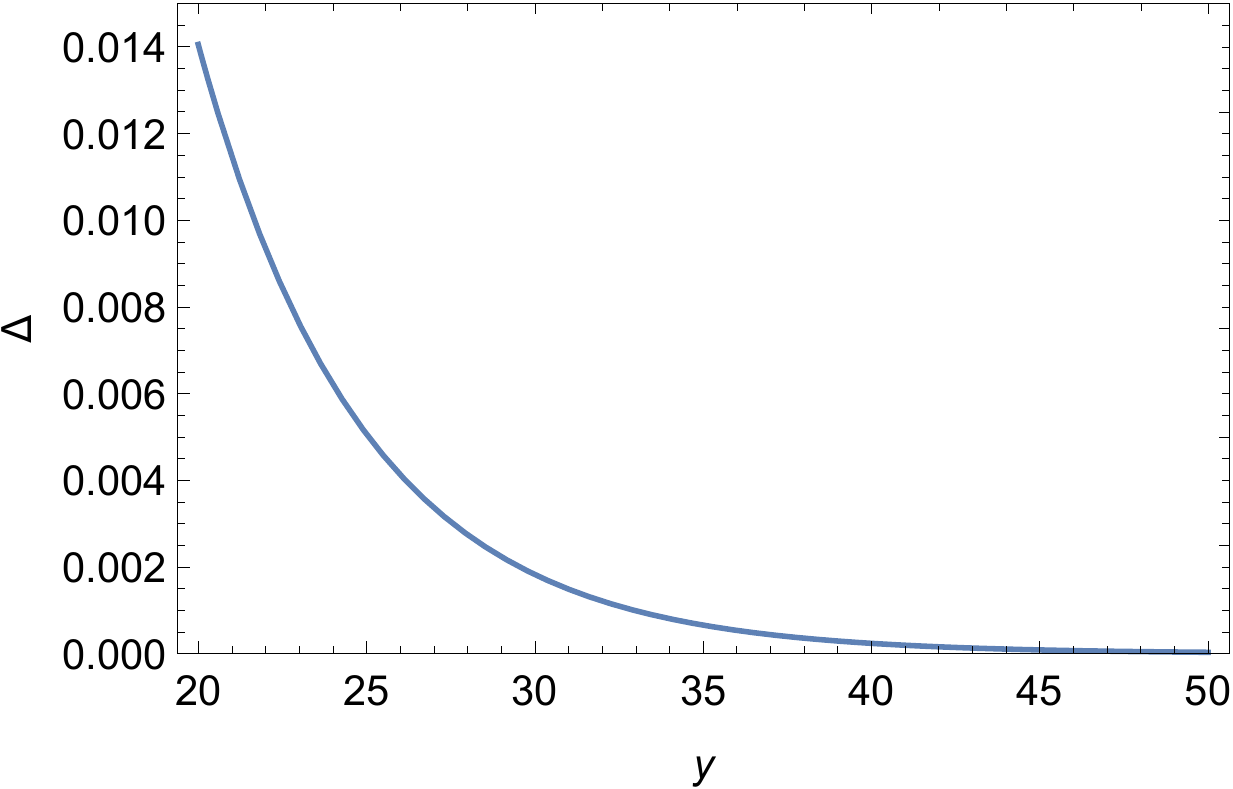}
	\caption{\label{onete}(Color online)The width dependence of the energy gap $ \Delta $ of the hinge states. The parameters are  $ m=2$, $ t=1$ , $ A=1$, $\varLambda=0.2 $.}
\end{figure}

Using a trial function $ \psi(y)=\psi_{\lambda}e^{\lambda y} $, the secular equation gives four roots of $ \lambda(E) $, denoted as $ n\lambda_{\kappa} $, with $ n\in{(+,-)}$,$ \kappa\in{(1,2)} $ and
\begin{equation}
	\lambda_{\kappa}^{2}(E)=-\dfrac{2F}{\varLambda^{2}}+(-1)^{\kappa-1}\dfrac{2\sqrt{F^{2}-\varLambda^{2}(D^{2}+A^{2}k_{z}^{2}-E^{2})}}{\varLambda^{2}}
\end{equation}
where $ F=-D\varLambda-A^{2} $ and $ D=\frac{\varLambda(m-3t)}{t} $. With the open boundary conditions of $ \Psi(k_{z},y=\pm\frac{L}{2})=0 $, we have an analytical expression for the wave function $ \Psi $
\begin{equation}
	\begin{aligned}
		\Psi_{1}=\widetilde{a}_{+}f_{+}(k_{z},y)+\widetilde{a}_{-}f_{-}(k_{z},y)\\
		\Psi_{2}=\widetilde{b}_{+}f_{+}(k_{z},y)+\widetilde{b}_{-}f_{-}(k_{z},y)\\
		\Psi_{3}=\widetilde{c}_{+}f_{+}(k_{z},y)+\widetilde{c}_{-}f_{-}(k_{z},y)\\
		\Psi_{4}=\widetilde{d}_{+}f_{+}(k_{z},y)+\widetilde{d}_{-}f_{-}(k_{z},y)
	\end{aligned}
\end{equation}
where we have difined
\begin{equation}
	\begin{aligned}
		f_{+}(k_{z},y)=\dfrac{\cosh(\lambda_{1}y)}{\cosh(\lambda_{1}\frac{L}{2})}-\frac{\cosh(\lambda_{2}y)}{\cosh(\lambda_{2}\frac{L}{2})}\\
		f_{-}(k_{z},y)=\dfrac{\sinh(\lambda_{1}y)}{\sinh(\lambda_{1}\frac{L}{2})}-\frac{\sinh(\lambda_{2}y)}{\sinh(\lambda_{2}\frac{L}{2})}
	\end{aligned}
\end{equation}
The nontrivial solution for the coefficients in the wave functions leads to a secular equation
\begin{equation}
	\dfrac{\tanh\frac{\lambda_{1}L}{2}}{\tanh\frac{\lambda_{2}L}{2}}+\frac{\tanh\frac{\lambda_{2}L}{2}}{\tanh\frac{\lambda_{1}L}{2}}=\dfrac{\xi_{1}^{2}\lambda_{2}^{2}+\xi_{2}^{2}\lambda_{1}^{2}+(\xi_{1}-\xi_{2})^{2}(\frac{E^{2}}{A^{2}}-k_{z}^{2})}{\xi_{1}\xi_{2}\lambda_{1}\lambda_{2}}
\end{equation}
 where $\xi_{1,2}=D-\frac{\varLambda}{2}\lambda_{1,2}^{2}$. The property of the solution of $ \lambda_{1,2} $ determines the spatial distribution of the wave function\cite{19}. In the large L limit, when $ \lambda_{1,2} $ is a pure imaginary number, the dispersion relation is $ E_{\pm}=\pm\frac{1}{2}\sqrt{(2D-\varLambda k_{y}^{2})^{2}+4A^{2}(k_{y}^{2}+k_{z}^{2})} $, i.e., the dispersion relation of the surface state; there exists a real $ \lambda_{1,2} $ when
 \begin{equation}
 	A^{2}>-\dfrac{2\varLambda^{2}(m-3t)}{t}>0
 \end{equation}
\begin{figure*}[htbp]
	\begin{minipage}[t]{0.45\textwidth}
		\includegraphics[width=8cm]{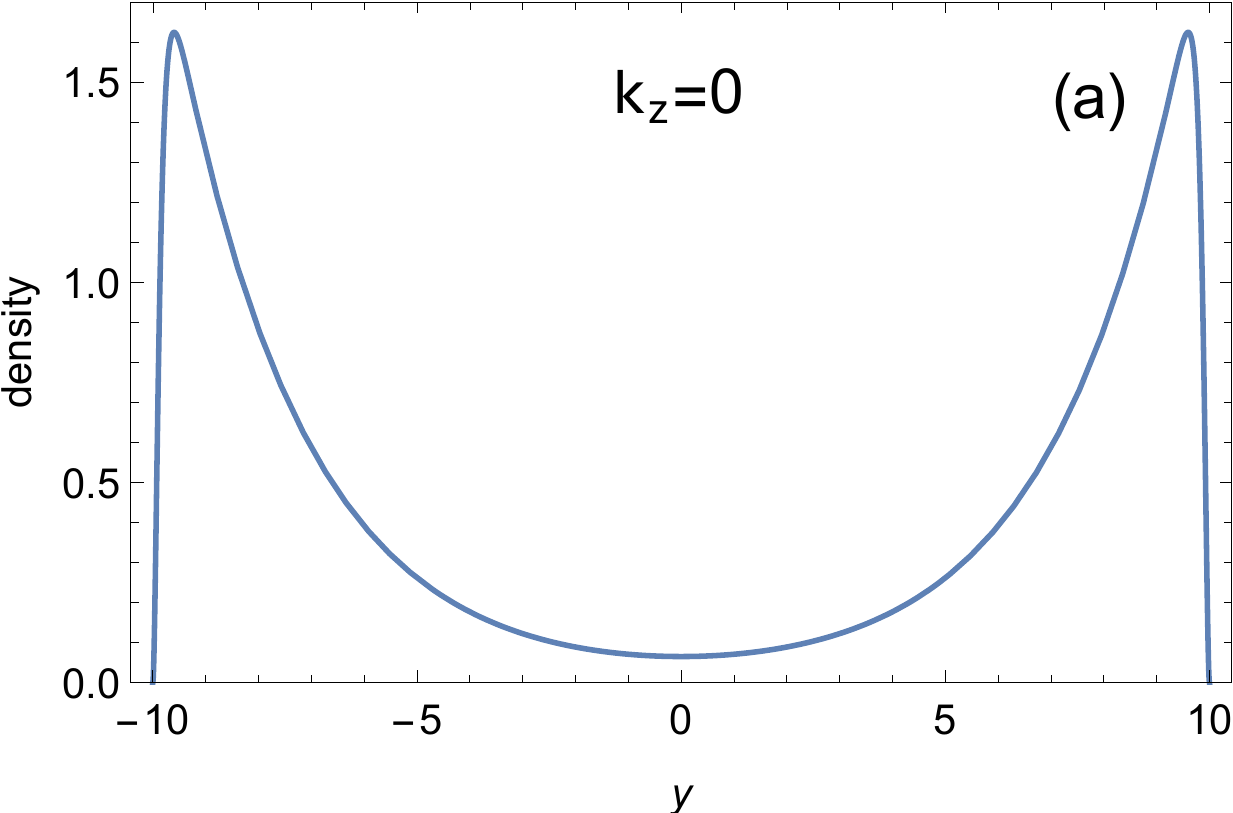}
	\end{minipage}
	\begin{minipage}[t]{0.45\textwidth}
		\includegraphics[width=8cm]{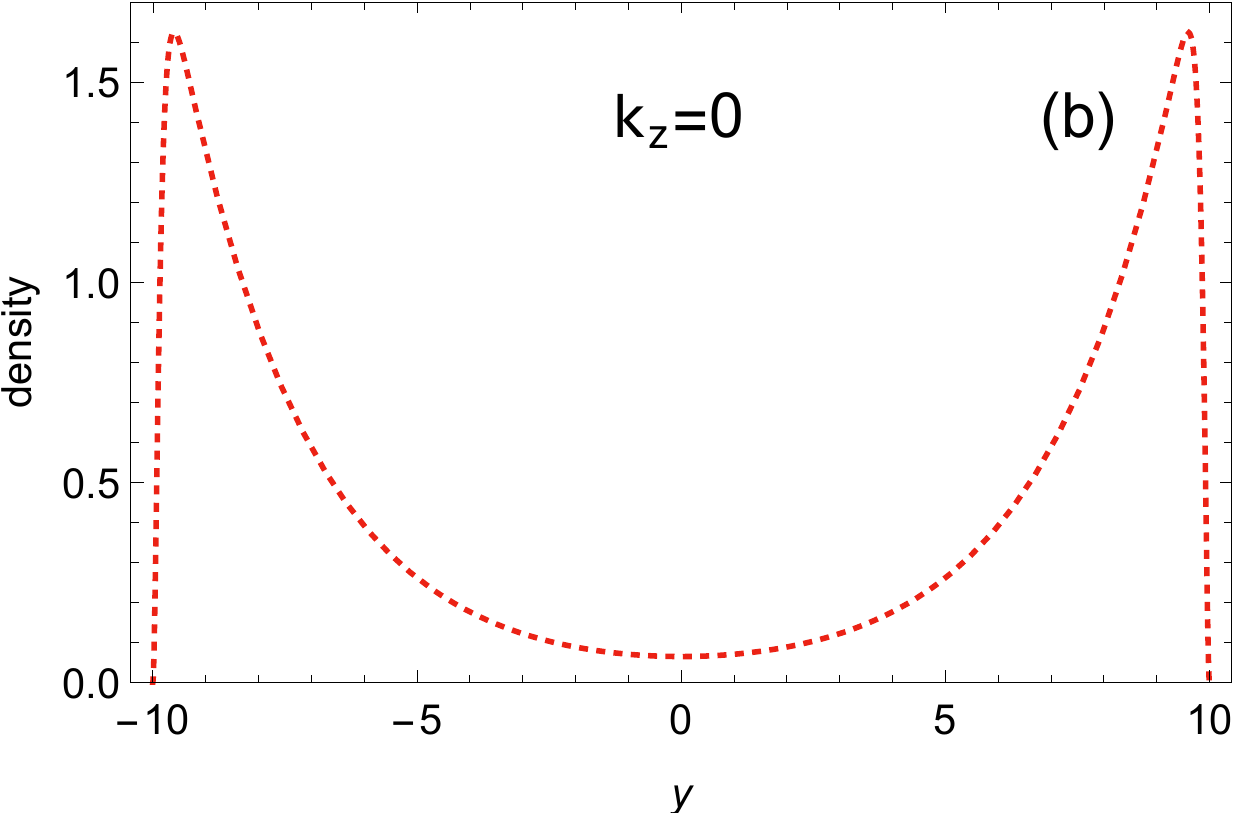}
	\end{minipage}
	\caption{\label{fig:2}(Color online)The density distribution of the two hinge states $ \Psi_{\pm}(k_{z},y) $ for $ L=20 $. (a) The solid line corresponds to $|\Psi_{+}(k_{z},y)|^{2}$; (b) The dotted line corresponds to $|\Psi_{-}(k_{z},y)|^{2}$.}
\end{figure*}
In the large L limit, near $ k_{z}=0 $, we have
\begin{equation}
	E_{\pm}(k_{z})=\pm Ak_{z}
\end{equation}
Our result is consistent with that of equation (16) in the previous section.\\

Due to the finite size effect, a finite energy gap can be opened because of coupling between the two hinge states. We define the gap as $ \Delta=E_{+}-E_{-} $ at the $ \Gamma $ point, where $ E_{+} $ and $ E_{-} $ are two solutions of equation (20). For real $ \lambda $ and finite $ L $, if $ \lambda_{1,2}L\gg1 $ and $ \lambda_{1}\gg\lambda_{2} $, 
\begin{equation}
	\Delta\backsimeq\dfrac{4|AD|}{\sqrt{A^{2}+2\varLambda D}}e^{-\lambda_{2}L}
\end{equation}
which decays exponentially with $L$. In Fig.\ref{onete}, we plot the gap function with $ 20\leq L\leq 50$.
For real $ \lambda  $s, the dispersion relation and gap of the 3DSOTI have different numerical solutions with different sizes. For simplicity, we take the parameter  $ m=2$, $ t=1$, $ A=1$, $\varLambda=0.2 $ for all numerical calculations in the paper. For $ L=50 $, the
energy gap is really tiny, $ \Delta= 3.07\times10^{-5} $. However, for $ L=20 $, the gap $ \Delta $ is $ 0.014 $, which becomes larger enough.\\

The expressions for the wave functions of the hinge states are
\begin{equation}
	\begin{aligned}
					\Psi_{+}=C_{+}\begin{pmatrix}
			\widetilde{a}_{+}^{+}f_{+}+\widetilde{a}_{-}^{+}f_{-}\\\widetilde{b}_{+}^{+}f_{+}+\widetilde{b}_{-}^{+}f_{-}\\\widetilde{c}_{+}^{+}f_{+}+\widetilde{c}_{-}^{+}f_{-}\\\widetilde{d}_{+}^{+}f_{+}+\widetilde{d}_{-}^{+}f_{-}
		\end{pmatrix};
		\Psi_{-}=C_{-}\begin{pmatrix}
			\widetilde{a}_{+}^{-}f_{+}+\widetilde{a}_{-}^{-}f_{-}\\\widetilde{b}_{+}^{-}f_{+}+\widetilde{b}_{-}^{-}f_{-}\\\widetilde{c}_{+}^{-}f_{+}+\widetilde{c}_{-}^{-}f_{-}\\\widetilde{d}_{+}^{-}f_{+}+\widetilde{d}_{-}^{-}f_{-}
		\end{pmatrix}	
	\end{aligned}
\end{equation}
with $ C_{\pm} $ are the normalization constants. The density distribution at the two hinges is mainly determined by the larger one of $ \lambda_{1,2}^{-1} $, according to the present analytic solutions of wave functions. In the example in Fig.\ref{onete}, we notice that  $ \lambda_{2}^{-1}\gg\lambda_{1}^{-1} $. For a larger size of the sample, the density of the wave function nearly vanishes far away from the edges if the width of the sample is considerably larger than the longer scale $ \lambda_{2}^{-1} $. For example, the density distributions of $ \Psi_{\pm} (k_{z},y)$ for $ L=20 $ are plotted for demonstration in Fig.\ref{fig:2} where $ \lambda_{2}^{-1} $ at $ k_{z}=0 $ is $ 4.8979 $, respectively. The densities of the wave functions $ \Psi_{\pm}(k_{z}=0,y) $ are symmetrically distributed at the two sides. This fact is consistent with the opening of an energy gap in the spectra at $ k_{z}=0 $.\\

\section{Conclusion}
In this paper, we show that 3DSOTI can generate energy gaps due to their limited size. Starting from the effective Hamiltonian of 3DSOTI, we obtain the effective Hamiltonian for the four 2D surface states with gaps by perturbative methods. Our results show that the mass term has alternating signs at the four hinges of the adjacent surfaces. Therefore, the four hinges of 3DSOTI have 1D gapless hinge states. We then obtain the effective Hamiltonian of the 1D gapless hinge state by perturbation calculation of the two-dimensional surface state with gaps. Through the Hamiltonian of a 2D surface state with finite width, we find that the hinge state on the two sides can be coupled to open a finite energy gap. Finally, we obtain the relation between the gap and the facets and show that an energy gap opens in the spectrum based on the wave function of the hinge state.

\begin{acknowledgments}
This work is supported by the National Natural Science Foundation of China (Grant No. No. 11904308).
\end{acknowledgments}

\end{document}